\documentstyle[11pt]{article}

\begin{document}

\begin{center}

{\Large {\bf Nuclear shell-model calculations for $^6\mbox{Li}$ and
$^{14}\mbox{N}$ with different NN potentials}}

\vspace{0.3in}

D. C. Zheng and B. R. Barrett\\
\begin{small}
{\it Department of Physics, University of Arizona, Tucson, AZ 85721}
\end{small}

\vspace{0.4in}

\end{center}

\thispagestyle{empty}

\vspace{0.2in}

\begin{abstract}
Two ``phase-shift equivalent'' local NN potentials with different
parametrizations, Reid93 and NijmII, which were found to give
nearly identical results for the triton by Friar {\it et al.},
are shown to yield remarkably similar results for
$^6\mbox{Li}$ and $^{14}\mbox{N}$ in a $(0+2)\hbar\Omega$
no-core space shell-model calculation.
The results are compared with those for the
widely used Hamada-Johnson hard-core and the original
Reid soft-core potentials, which have
larger deuteron $D$-state percentages.
The strong correlation between the tensor strength and the
nuclear binding energy is confirmed.
However, many nuclear-structure properties seem to be
rather insensitive to the details of the NN potential and, therefore,
cannot be used to test various NN potentials.
\end{abstract}

\pagebreak

\section{Introduction}
Currently a large variety of nucleon-nucleon (NN) potentials
exist. Despite the fact that all these potentials
were designed to more or less reproduce the deuteron
properties and the low-energy NN scattering data, they are quite
different in details. In particular, the deuteron $D$-state
percentage given by different potentials ranges from
4.4\% for Bonn A \cite{bonn} to 7.0\% for Hamada-Johnson \cite{hj}.
Exact (or nearly exact) calculations for the $A$=3 and $A$=4
systems \cite{friar,glockle}
show that there is a strong correlation between the
tensor strength of the one-boson-exchange NN potential and the binding energy.
The NN potential which has a weaker tensor strength tends to
yield a larger binding energy.

In Ref.\cite{friar}, however, it is found that when
the NN potentials with different parametrizations are optimally
fitted (with $\chi^2_{\rm min}$ per datum about 1)
to the phase-shifts obtained in the ``new and
comprehensive (energy-dependent) partial-wave analysis'',
they result in nearly identical triton properties. It is pointed out
that the tensor force is tightly constrained by the mixing parameter
$\epsilon_1$, which can be determined very well in their multi-energy
analysis. For example, three local NN potentials, NijmII, Reid93 and AV18
with parameters newly determined by their fit, have deuteron $D$-state
percentages of 5.64\%, 5.70\% and 5.65\% respectively.

It is the main purpose of this work to find out whether or not the above
observation made in the study of the $A$=3 system also holds
for larger nuclei. To this end, we will perform nuclear
shell-model (SM) calculations for $^6\mbox{Li}$ and $^{14}\mbox{N}$
using effective interactions derived from the Reid93 and NijmII
potentials. To set a scale for the comparison, the same calculations will
be performed using the Hamada-Johnson hard-core potential (H-J) \cite{hj} and
the original Reid-soft-core potential (RSC) \cite{reid}, both of which
were widely used in nuclear-structure calculations in the past.
We will adopt a no-core model space \cite{nocore} for which the effective
interaction can reasonably be approximated by the Brueckner G-matrix
\cite{bruc} evaluated at appropriate starting energies. This
avoids the difficulty encountered in doing the infinite summation
over the core-polarization diagrams for
the effective interaction in the presence of an inert core.
Of course, for each nucleus,
the effective interactions will be calculated in exactly the same manner
so the difference reflected in the nuclear SM results
is solely due to the difference in the NN potentials.

\section{Calculations}
The Brueckner G-matrix \cite{bruc} is calculated according to the following
equation using the technique developed in Ref.\cite{bhm}:
\begin{equation}
G(\omega) = v_{12} + v_{12}\frac{Q}{\omega - (h_1+h_2+v_{12})} v_{12},
		\label{g}
\end{equation}
where $\omega$ is the starting energy,
$v_{12}$ is the NN potential, $h$ is the single-particle Hamiltonian
with the one-body mean field ${\cal V}$ approximated by
a shifted harmonic-oscillator (HO)
potential $u$:
\begin{equation}
h = t+ {\cal V} \simeq t + u - V_0
	= \frac{p^2}{2m} + \frac{1}{2}m\Omega^2 r^2 - V_0.
\end{equation}
The parameter $V_0$ approximately represents the depth of the
mean field ${\cal V}$ in which the two nucleons in the ladder-scattering
processes are moving. This parameter can be absorbed into the
starting energy by writing the G-matrix equation (\ref{g}) as
\begin{equation}
G(\omega) = v_{12} + v_{12}
\frac{Q}{\omega' - (h^{\rm HO}_1+h^{\rm HO}_2+v_{12})} v_{12}, \label{g'}
\end{equation}
where $\omega'=\omega+2V_0$
and $h^{\rm HO} = \frac{p^2}{2m} + \frac{1}{2}m\Omega^2 r^2$.

For the two-body G-matrix element $\langle ab|G|cd\rangle_{J,T}$,
the shifted starting energy $\omega'$ is taken as
\begin{equation}
\omega' = \epsilon_a + \epsilon_b + \Delta ,      \label{sw}
\end{equation}
where $\epsilon$ are the HO single-particle energies:
\begin{equation}
\epsilon_a = \left(2n_a+l_a+\frac{3}{2}\right) \hbar\Omega.
\end{equation}
Roughly speaking, the quantity $\Delta$ represents the interaction
energy of the
two nucleons bound by the HO potential. It is negative and depends on
the two-nucleon state. In this work, $\Delta$ is regarded as
an adjustable parameter whose value will be so chosen that
we obtain a reasonable binding energy. It should be emphasized that
$\Delta$ will only be adjusted when we go from one nucleus to another;
its value is fixed when we consider the same nucleus but
different NN potentials. We chose $\hbar\Omega=16$ MeV, $\Delta=-38$ MeV
for $A$=6 and $\hbar\Omega=14$ MeV, $\Delta=-62$ MeV for $A$=14.

In the no-core model space, the starting-energy independent
two-body effective interaction is simply the sum of ladder diagrams (G-matrix)
plus the folded diagrams. When the starting energy for the
G-matrix is properly chosen, the folded diagrams have a small contribution
to the effective interaction, which can, therefore, be
well approximated by the G-matrix alone.

Since we will be performing a $(0+2)\hbar\Omega$ calculation,
the Pauli operator $Q$ is defined as ($N_1=2n_1+l_1$ and
$N_2=2n_2+l_2$ are the principal quantum numbers for the
two single-particle orbitals of the intermediate states in
the ladder diagrams):
\begin{eqnarray}
{\rm For}\; ^6\mbox{Li}: \hspace{0.2in}
Q &=& 1  \hspace{0.15in} {\rm for}\hspace{0.05in}
	(N_1+N_2) > 4, N_1\neq 0, \hspace{0.05in} {\rm and}
	\hspace{0.05in} N_2\neq 0; \nonumber \\
  &=& 0 \hspace{0.15in} {\rm for}\hspace{0.05in}
	(N_1+N_2) \le 4, N_1=0, \hspace{0.05in} {\rm or}
	\hspace{0.05in} N_2= 0. \\
{\rm For}\; ^{14}\mbox{N}: \hspace{0.2in}
Q &=& 1  \hspace{0.15in} {\rm for}\hspace{0.05in}
	(N_1+N_2) > 4, N_1 > 1, \hspace{0.05in} {\rm and}
	\hspace{0.05in} N_2 > 1; \nonumber \\
  &=& 0 \hspace{0.15in} {\rm for}\hspace{0.05in}
	(N_1+N_2) \le 4, N_1 \le 1, \hspace{0.05in} {\rm or}
	\hspace{0.05in} N_2 \le 1.
\end{eqnarray}
Namely, in the ladder diagrams, the scattering into a two-particle
state with $N_1+N_2 \le 4$ is forbidden because
these states will be included in the $(0+2)\hbar\Omega$ calculation.
In addition, for $^6\mbox{Li}$, the scattering into a two-particle state with
one particle in the $0s$ orbit is also forbidden because the
$0s$ orbit is mostly occupied.
For $^{14}\mbox{N}$, the scattering into a two-particle state with
one particle in the $0s$ or $0p$ orbits is not allowed for
the same reason. This is also shown in Fig.1. Notice that in Fig.1
we have cut off the ``wings'' which extend
out from the mostly occupied orbitals, at
a large enough $N$, so as to obtain a converged result.
It should be pointed out that
our treatment of the Pauli operator $Q$ is not exact;
a more accurate definition is possible.
Such a definition should be reasonable enough to allow us
to do sensible calculations and meaningful comparisons. Here a more
important issue is that the effective interactions for different
potentials are calculated in exactly the same way.

Once the G matrices are obtained, the Hamiltonian for the SM
calculation is written as:
\begin{equation}
H_{\rm SM} = \left(\sum_{i=1}^A t_i -T_{\rm CM}\right) + \sum_{i<j}^A G_{ij}
		+ \lambda \left(H_{\rm CM}-\frac{3}{2}\hbar\Omega\right),
\end{equation}
where $T_{\rm CM}$ is the center-of-mass (CM) kinetic energy and
$H_{\rm CM}$ is the HO CM Hamiltonian. The parameter $\lambda$ is taken
to be large enough ($\sim 10$) to force the CM motion into its lowest
configuration for all the low-lying states.

\section{Results and Discussions}
Our calculations are performed using the OXBASH SM code \cite{oxbash}.
The results for the ground-state energy, the low-lying positive-parity
energy spectrum, and selected electromagnetic (EM) static and transitional
properties in $^6\mbox{Li}$ and $^{14}\mbox{N}$ are shown in Table I for
the four NN potentials used in this work. The deuteron $D$-state
percentage $P_D$ for each potential is also listed.

Before we proceed to discuss the results,
it should be emphasized that our calculations involve some
parameters, such as $\Delta$ and $\hbar\Omega$.
The calculated ground-state energy, which is quite
sensitive to the choice of these parameters, should not be taken as
an accurate result. Nevertheless, since we perform exactly the
same calculation for different potentials, the results are valid
for the purpose of comparisons.

In Ref.\cite{friar}, the Reid93 and NijmII potentials have been shown
to give almost the same results for the triton by solving the Faddeev equation
\cite{faddeev}. It is evident from Table I that
the effective interactions calculated
from these potentials also produce remarkably similar results for the more
complex $A$=6 and 14 systems. This is more obvious when we compare the
results for these two potentials with those for the H-J and RSC potentials.
Note that the similarity between the Reid93 and NijmII results lies not only
in the bulk properties, such as the binding energy, but also in more subtle
observables, like excitation energies and EM transition rates.
The latter reflects the equivalence in the wave functions for these
two interactions. It is then safe to conclude that
the Reid93 and NijmII potentials are not only
equivalent in describing the $A$=2 and 3 systems as shown in Ref.\cite{friar},
they are also equivalent for the larger systems
$^6\mbox{Li}$ and $^{14}\mbox{N}$ studied in this work.

For the $A$=3 and $A$=4 systems, it has been found that there is a
strong correlation between the nuclear binding energy and
the strength of the tensor component in the NN potential,
characterized by the deuteron $D$-state percentage $P_D$ \cite{friar,glockle}.
Apparently, the above correlation persists for the more complex nuclei
$^6\mbox{Li}$ and $^{14}\mbox{N}$ in a truncated-space SM calculation,
as performed here. The calculated binding energies
for the HJ and RSC potentials, which have larger deuteron $D$-state
percentages, are much smaller than those
for the Reid93 and NijmII potentials.

Note, however, that not all nuclear-structure properties are sensitive
to the details of the NN potential. By examining Table I,
one can see that the {\em excitation} energies for the low-lying states
and some EM observables for different NN potentials are actually
quite similar. This makes it difficult to
test the NN potential in a nucleus by studying nuclear structure.

On the other hand, besides the binding energy,
there do exist a few observables which
have a strong dependence on one or more particular components of the NN
potential. Two famous examples are the isospin splitting between the
energies of the $J^{\pi}=0^-_1$, $T$=0 and
$J^{\pi}=0^-_1$, $T$=1 states in $^{16}\mbox{O}$
and the Gamow-Teller strength $B({\rm GT})$ between the ground state in
$^{14}\mbox{C}$ and $^{14}\mbox{N}$. The former is sensitive to the
tensor component in the NN potential \cite{o16} and the latter is sensitive
to the spin-orbit and tensor components \cite{c14}.
These have been studied in great detail
in \cite{zz}. In Table I, one does see a sizable change in the
$B({\rm GT})$ value from the ground state of $^{14}\mbox{C}$,
when going from one potential to the other. The $B({\rm M1})$
value is closely related to the $B({\rm GT})$ value and also
shows some change. However, both the $B({\rm GT})$ and $B({\rm M1})$
values are not a monotonic function of $P_D$, suggesting that
besides the tensor force strength, other differences between the
four NN potentials also play a role.

{}From Table I, we see that the calculated quadrupole moment for the
ground state of $^6\mbox{Li}$ also experiences quite significant changes
when the NN potential is altered. In our truncated-space SM calculation
with bare operators [i.e., $e(p)=1$ and $e(n)=0$], none of the NN potentials
yields the experimental result. The smallness of the
experimental quadrupole moment for $^6\mbox{Li}$ and
of the $B({\rm GT})$ value from $^{14}\mbox{C}$ seems
to indicate that in a truncated SM calculation,
one needs a weaker tensor force. This is consistent with the observations
made in Ref.\cite{zz}, where some other quantities
which are sensitive to the tensor force were considered.

\section{Conclusions}
In this work, we perform a no-core space $(0+2)\hbar\Omega$
SM calculation for the $A$=6 and 14 systems
using effective interactions calculated from
four different NN potentials.
The Reid93 and NijmII potentials, which were found to be equivalent
in the description of the $A$=2 and $A$=3 systems, are shown to yield almost
identical results, not only for the binding energy, but also for the
energy spectrum and the EM static and transitional properties
for the larger nuclei considered in this work.
It is verified that the NN potential with a weaker
tensor component (i.e., smaller deuteron $D$-state percentage)
tends to give a larger nuclear binding energy, consistent
with the observations made in the study of the smaller $A$=3 and 4 systems.
Many nuclear-structure properties are found to be rather insensitive
to the details of a NN potential and similar results are obtained for
them, when different NN potentials are used.

Among the various observables considered, there are a few which are
sensitive to the various components of a NN potential, in particular,
the tensor force.
In order to reproduce the experimental results for the GT transition
strength from the ground state of $^{14}\mbox{C}$ and the
quadrupole moment of $^6\mbox{Li}$ in a truncated space SM calculation,
a weaker tensor force is needed.

\section*{Acknowledgment}
We thank J. J. de Swart and M. Rentmeester for providing us the
NN potentials Reid93 and NijmII. This work was supported by
the National Science Foundation, Grant No. PHY-9103011.

\pagebreak

\begin{small}

\noindent
{\bf Table I.}  The ground-state energies and excitation
energies of low-lying states, in units of MeV,
of $^6\mbox{Li}$ and $^{14}\mbox{N}$ in a $(0+2)\hbar\Omega$
calculation.
The G-matrices are calculated in exactly the same way
from different NN potentials including
Hamada-Johnson (H-J), Reid-soft-core (RSC), new Reid-soft-core (Reid93),
and new Nijmegen (NijmII).
For $^6\mbox{Li}$ ($^{14}\mbox{N}$), a harmonic-oscillator
basis with $\hbar\Omega=16$ MeV ($\hbar\Omega=14$ MeV) is used.
The GT strength $B({\rm GT})$ and M1 strength $B({\rm M1})$ (in $\mu^2_N$) is
calculated for the transition $0^+_1(1)\rightarrow 1^+_1(0)$.
The magnetic dipole moment $\mu$ (in $\mu_N$) and
electric quadrupole moment $Q$ (in $e {\rm fm}^2$) are calculated for
the $J^{\pi}(T)=1^+_1(0)$ state. Bare operators
[$g_l(p)=1$, $g_l(n)=0$, $g_s(p)=5.582$, $g_s(n)=-3.826, e(p)=1, e(n)=0$]
are used in these calculations.
In the Table, $P_D$ is the
deuteron $D$-state percentage.

\begin{center}
\begin{tabular}{c|c|rrrr|c}\hline\hline
Nucleus & $J^{\pi} (T)$ & H-J & RSC & Reid93 & NijmII & Expt.\\ \hline
$^2\mbox{H}$  & $P_D$   & 7.00\% & 6.47\% & 5.70\% & 5.64\% & N/A \\ \hline
$^6\mbox{Li}$ & $1^+_1 (0)$ &-21.353 &-22.545 &-24.997 &-26.038 &-31.996\\
              & $3^+_1 (0)$ &  3.271 &  2.946 &  2.822 &  2.840 &  2.186\\
              & $0^+_1 (1)$ &  3.977 &  3.691 &  3.605 &  3.566 &  3.563\\
              & $2^+_1 (0)$ &  5.275 &  5.038 &  5.400 &  5.448 &  4.310\\
              & $2^+_1 (1)$ &  7.197 &  6.816 &  6.898 &  6.906 &  5.366\\
              & $1^+_2 (0)$ &  7.592 &  7.617 &  8.101 &  8.205 &  5.650\\
              & $2^+_2 (1)$ & 11.854 & 11.530 & 11.842 & 11.957 &       \\
              & $1^+_1 (1)$ & 12.840 & 12.586 & 13.065 & 13.236 &       \\
              & $0^+_2 (1)$ & 16.307 & 16.345 & 17.001 & 17.204 &       \\
            & $B({\rm GT})$ &  5.327 &  5.335 &  5.387 &  5.397 &       \\
            & $B({\rm M1})$ &  15.21 &  15.38 &  15.53 &  15.57 &       \\
              & $\mu$       &  0.840 &  0.842 &  0.848 &  0.849 &  0.822 \\
              & $Q$         & -0.265 & -0.208 & -0.168 & -0.163 & -0.11  \\
								\hline
$^{14}\mbox{N}$
              & $1^+_1 (0)$ &-56.306 &-75.944 &-87.562 &-87.087 & -104.659  \\
              & $0^+_1 (1)$ &  2.176 &  1.894 &  1.702 &  1.637 &   2.313 \\
              & $1^+_2 (0)$ &  2.590 &  2.196 &  2.043 &  2.033 &   3.948 \\
              & $2^+_1 (0)$ &  3.587 &  3.374 &  3.615 &  3.594 &   7.190 \\
              & $2^+_1 (1)$ &  6.434 &  6.660 &  6.874 &  6.801 &  10.149 \\
              & $3^+_1 (0)$ &  8.349 &  9.227 &  9.779 &  9.757 &  11.050\\
              & $1^+_1 (1)$ &  9.370 &  9.492 &  9.971 &  9.971 &  13.619 \\
             & $B({\rm GT})$&  1.953 &  1.243 &  1.297 &  1.305 & 0     \\
             & $B({\rm M1})$&  5.380 &  3.694 &  3.823 &  3.842 &       \\
              & $\mu$       &  0.562 &  0.495 &  0.499 &  0.499 &       \\
              & $Q$         &  2.285 &  2.155 &  2.105 &  2.109 &      \\
								\hline\hline
\end{tabular}
\end{center}

\end{small}

\pagebreak

\section*{Figure Caption}

{\bf Fig.1} The Pauli operator $Q$ for $A$=6 and $A$=14 used in this work.


\vspace{0.2in}

\begin{thebibliography}{99}
\bibitem{bonn} R. Machleidt, Adv. in Nucl. Phys. Vol. 19, 189(1989).
\bibitem{hj} T. Hamada and I.D. Johnson, Nucl. Phys. {\bf 34}, 382(1962).
\bibitem{friar} J.L. Friar, G.L. Payne, V.G.J. Stoks, and J.J. de Swart,
	Phys. Lett. {\bf B311}, 4(1993).
\bibitem{glockle} W. Gl\"ockle and H. Kamada,
	Phys. Rev. Lett. {\bf 71}, 971 (1993).
\bibitem{reid} R.V. Reid, Ann. of Phys. {\bf 50}, 411 (1968).
\bibitem{nocore} J.P. Vary, in {\it Theory and Applications of Moment
	Methods in Many-Fermion Systems}, eds B.J. Dalton, S.M. Grimes,
	J.P. Vary, and S.A. Williams, Plenum Press (New York), 1980, p.423.
\bibitem{bruc} K.A. Brueckner, Phys. Rev. {\bf 97}, 1353(1955);
		 {\bf 100}, 36(1955).
\bibitem{bhm} B.R. Barrett, R.G.L. Hewitt, and R.J. McCarthy,
	Phys. Rev. {\bf C 3}, 1137(1971).
\bibitem{oxbash} A. Etchegoyen, W.D.M. Rae, N.S. Godwin, B.A. Brown,
	W.E. Ormand, and J.S. Winfield, the Oxford--Buenos Aires --- MSU
	Shell Model Code (OXBASH) (unpublished).
\bibitem{faddeev} L.D. Faddeev, Zh. Eksp. Theo. Fiz. {\bf 39}, 1459(1960).
\bibitem{o16} J. Blomqvist and A. Molinari, Nucl. Phys.
        {\bf A106}, 545(1968); D.J. Millener and D. Kurath, Nucl. Phys.
        {\bf A255}, 315(1975); B.R. Barrett, Phys. Rev. {\bf 159}, 816(1967).
\bibitem{c14} C.W. Wong, Nucl. Phys. {\bf A108}, 481(1968);
	R.R. Scheerbaum, Phys. Lett. {\bf 63B}, 381(1976).
\bibitem{zz} D.C. Zheng and L. Zamick, Ann. Phys. (NY) {\bf 206}, 106(1991);
	L. Zamick, D.C. Zheng, and H. M\"{u}ther,
	Phys. Rev. {\bf C 45}, 2763(1992);
	L. Zamick and D.C. Zheng, Phys. Rep., in press.
\end{thebibliography}
\end{document}